\documentclass[aps,prl,twocolumn,superscriptaddress,showpacs]{revtex4-1}
\usepackage{graphicx}
\usepackage{dcolumn}
\usepackage{bm}
\usepackage{float}
\usepackage{amssymb}
\usepackage{amsmath}
\usepackage{ulem}

\usepackage{hyperref}
\usepackage[mathlines]{lineno}
\usepackage[dvipsnames]{xcolor}
\setpagewiselinenumbers
\modulolinenumbers[5]

\usepackage{soul}

\newcommand{\mdel}[1]{\ \unskip}
\newcommand{\del}[1]{\ \unskip}
\newcommand{\add}[1]{\textcolor{black}{#1}}

\usepackage{amsthm}
\usepackage{mathtools}
\usepackage{physics}
\usepackage{xcolor}
\usepackage{graphicx}
\usepackage[left=23mm,right=13mm,top=35mm,columnsep=15pt]{geometry} 
\usepackage{adjustbox}
\usepackage{placeins}
\usepackage[T1]{fontenc}
\usepackage{lipsum}
\usepackage{csquotes}
\begin{document}
\title{Collisional loss of one-dimensional fermions near a $p$-wave Feshbach resonance}


\author{Ya-Ting Chang}
    \affiliation{Department of Physics and Astronomy, Rice University, Houston, Texas 77005, USA}

\author{Ruwan Senaratne}
    \affiliation{Department of Physics and Astronomy, Rice University, Houston, Texas 77005, USA}
    
\author{Danyel Cavazos-Cavazos}
    \affiliation{Department of Physics and Astronomy, Rice University, Houston, Texas 77005, USA}
    
\author{Randall G. Hulet}
    \email[ ]{randy@rice.edu}
    \affiliation{Department of Physics and Astronomy, Rice University, Houston, Texas 77005, USA}    
\date{\today} 

\begin{abstract}
We study collisional loss of a quasi-one-dimensional (1D) spin-polarized Fermi gas near a $p$-wave Feshbach resonance in ultracold $^6$Li atoms. We measure the location of the $p$-wave resonance in quasi-1D and observe a confinement-induced shift and broadening. We find that the three-body loss coefficient $L_3$ as a function of the quasi-1D confinement has little dependence on confinement strength. We also analyze the atom loss with a two-step cascade three-body loss model in which weakly bound dimers are formed prior to their loss arising from atom-dimer collisions. Our data are consistent with this model. We also find a possible suppression in the rate of dimer relaxation with strong quasi-1D confinement. We discuss the implications of these measurements for observing $p$-wave pairing in quasi-1D.

\end{abstract}

\maketitle

The realization of ultracold atomic Fermi gases has provided experimental access to a wide array of phenomena, largely because of the presence of Feshbach resonances \add{(FRs)} that provide for externally tunable interactions \cite{DeMarco1999,Truscott2001,Schreck2001,Granade2002}. In addition to the usual $s$-wave interactions between distinguishable fermions, higher partial-wave interactions may be tuned via \del{Feshbach resonance}\add{FR}s \cite{Chin2010}. $p$-Wave interactions are of particular interest as they are the dominant low-energy scattering process between identical fermions and are predicted to exhibit phenomena distinct from those observed in $s$-wave interacting Fermi gases \cite{Gurarie2007}. In particular, pairing between identical fermions is an essential ingredient of the Kitaev chain Hamiltonian \cite{Kitaev2000}, which supports Majorana zero-modes at the ends of the chain. These zero-modes have been observed in semiconducting nanowires \cite{Sankar2015}, and are a promising candidate platform for fault-tolerant quantum computing \cite{Kitaev2002,Stern2013}.

$p$-Wave \del{Feshbach resonance}\add{FR}s have been observed in $^{40}$K \cite{Regal2003,Gunter2005,Thywissen2016} and $^6$Li \cite{Salomon2004,Ketterle2005,Mukaiyama2008,Vale2008,Zimmermann2010,Weidemuller2019}. The severe atom losses associated with these resonances, however, have limited their usefulness. Three-body losses, which are suppressed by symmetry in the case of a fermionic two-spin system with $s$-wave interactions \cite{Shlyapnikov2004}, are not suppressed for $p$-wave interactions. Much work has been done in characterizing the atom loss associated with $p$-wave \del{Feshbach resonance}\add{FR}s \cite{Mukaiyama2016,Mukaiyama2017,Mukaiyama2018a, Mukaiyama2018b}, and there is renewed interest in studying these resonances in reduced dimensions. Recent theoretical work has suggested that three-body losses may be suppressed in quasi-1D \cite{Cui2017}. The absence of a centrifugal barrier in 1D results in Feshbach dimers that have extended wavefunctions which overlap less with deeply-bound molecules. If three-body loss is suppressed by this mechanism, it might open a path towards realizing $p$-wave pairing in quasi-1D and emulating the Kitaev chain Hamiltonian.

We present an experimental study of three-body losses near a $p$-wave \del{Feshbach resonance}\add{FR} of identical  $^6$Li fermions in quasi-1D. We measure the three-body loss coefficient ($L_3$) as a function of 1D confinement for a direct three-body process. We also analyze the observed atom loss within the framework of a cascade model \del{, in which a transient Feshbach dimer formed in an initial atom-atom collision is subsequently lost due to the formation of a deeply-bound molecule in an inelastic atom-dimer collision}\add{with explicit dimer formation and relaxation steps} \cite{BoGao2018,Mukaiyama2019}, using \textit{in situ} imaging to reduce the effect of the inhomogeneous density. Finally, we characterize the confinement-induced shifts in the resonance position that appear in quasi-1D \cite{Olshanii98,Olshanii2003,Granger2004,Pricoupenko2008,Shlyapnikov2017}. These shifts allow us to extract a value for the effective range.

The apparatus and the experimental methods we use to prepare degenerate Fermi gases have been described previously \cite{Pedro2015,Ernie2018,Randy2020}. A $^6$Li degenerate Fermi gas is first prepared in the two lowest hyperfine states of the $S_{1/2}$ manifold (states $\ket{1}$ and $\ket{2}$, respectively) at 595 G, and then loaded into a crossed-beam dipole trap formed by three linearly-polarized mutually-orthogonal laser beams of wavelength $\lambda=1.064$ $\mu$m. Each beam is retro-reflected, with the polarizations of the incoming and retro-reflected beams \add{initially} set to be perpendicular to each other to avoid lattice formation. We eliminate state $|1 \rangle$ from the trap with a resonant burst of light. At this stage, we obtain $9(1)\times 10^4$ atoms in state $\ket{2}$ in a nearly isotropic harmonic trap with a geometric-mean trapping frequency of $2\pi\times 305(2)$ Hz, and at a temperature $T/T_F \approx 0.1$ where, $T_F$ is the Fermi temperature. The optical trap depths are increased and the polarizations of the retro-reflected beams are rotated to achieve a 7 $E_r$ deep 3D optical lattice, where $E_r=h^2 /(2m\lambda^2)=k_\mathrm{B}\times1.41$ $\mu \text{K}$ is the recoil energy, and $m$ is the atomic mass. During the lattice ramp\add{-up}, a co-propagating beam of 532 nm light is introduced along each trapping-beam dimension to flatten the trapping potential \cite{Pedro2015,Ernie2018}. By tuning these compensation beam powers, we create a 3D band insulator with a central density of approximately 1 atom per site. In order to produce a 2D lattice, which is an array of quasi-1D tubes, we slowly turn off the compensation beams and the vertical lattice beam, while increasing the intensity of the two remaining beams to achieve a desired 2D lattice depth, $V_L$. This depth determines the confinement in the quasi-1D traps, which is parameterized by $a_\perp=\sqrt{\add{2}\hbar/m\omega_r}$ \add{transversely} and $R_F$ \add{axially}, where $\omega_r=\sqrt{4 E_r V_L}/\hbar$ is the trapping frequency of a lattice site when approximated as a harmonic potential, and $R_F(N_{t,j}, \omega_z)=\sqrt{(2 N_{t,j}+1)\hbar/m \omega_z}$ is the Fermi radius of tube $j$ with number of atoms $N_{t,j}$ and an axial frequency $\omega_z$. The aspect ratio of the quasi-1D tubes, $\omega_r/\omega_z\approx\mdel{165}\add{170}$. We load a maximum of around 30 atoms per quasi-1D tube with $T<T_F$ to avoid exciting any radial modes.

We use a two-step servo scheme to stabilize the current in the coils producing the Feshbach magnetic field, because the $^6$Li $\ket{1}-\ket{1}$ $p$-wave \del{Feshbach resonance}\add{FR} near 159 G is very narrow. The first servo, $S_1$, provides the large dynamic range required to run our experimental sequence, while the second servo, $S_2$, controls the current in a bypass circuit added in parallel to the magnetic coils. This improves the stability of the magnetic field to  $\pm$10 mG and provides finer magnetic-field resolution. After reaching the hold field $B$, the atoms are transferred into $|1 \rangle$ with a $\pi$-pulse of duration 75 $\mu$s using RF radiation resonant with the $\ket{1}-\ket{2}$ transition. After a hold time $\tau$, we ramp the field back to 595 G, where the distribution of the remaining atoms is imaged using \textit{in situ} phase-contrast imaging \add{with a probe beam propagating perpendicular to the tube axis} \cite{Randy2020}. By using the inverse Abel transform, which exploits the approximate cylindrical symmetry of the 2D lattice, we measure the distribution with a spatial resolution of approximately three lattice constants. We sector the 2D lattice into concentric shells in which the tubes have similar chemical potentials\add{, }$\mu$\del{, as shown in Fig. 1}. This procedure is useful as scattering processes are in general energy-dependent, so observables depend on rate coefficients that are averaged over the Fermi-Dirac distribution for atoms in each tube.



\begin{figure}[t!]
    \centering
    \includegraphics[width=0.48\textwidth]{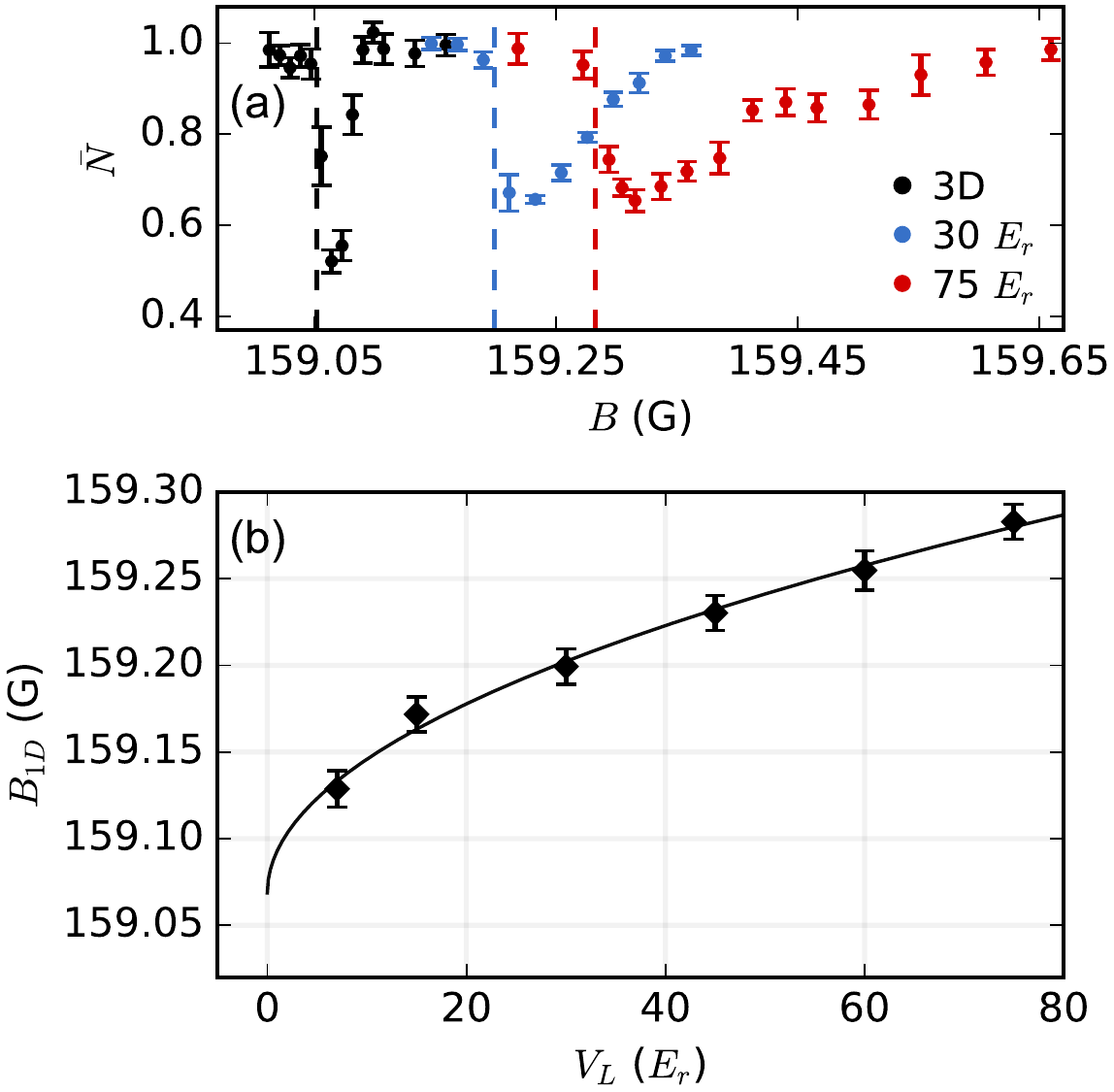}
    \caption{\del{FIG. 2}(a) $p$-Wave resonances in 3D and quasi-1D measured with magnetic-field-dependent loss.\del{ We define $\delta_B(V_L, \alpha_p)=B_{1D}-B_{3D}$, where we measure $B_{3D}=159.05(1)$ G. The resonance position for each case (dashed line) corresponds to the onset (15\% loss) of the observed atomic loss averaged over 6 experimental runs.} \add{Dashed lines show the resonance position for each $V_L$. We define the resonance field for zero-momentum collisions, which corresponds to the onset (15\% loss, to overcome atom number fluctuation) of the observed atomic loss. Data are averaged over 6 experimental runs and e}\del{E}rror bars are the standard error of the mean. (b) Diamonds show \del{$\delta_B$}\add{$B_{1D}$} \del{as a function of}\add{vs} $V_L$. \add{The} \del{S}\add{s}olid curve shows the result of fitting the data to Eq. \ref{eq:shift}, where the effective range $\alpha_p = $ 0.14(1) $a_0^{-1}$ and $B_{3D} = $ 159.07(1) G are fitted parameters. Error bars \del{indicate}\add{are} the statistical uncertainty arising from atom number fluctuation and field instability. \add{In both (a) and (b), $\tau$ is chosen such that peak loss is 30-50\% of total atom number for each value of $V_L$: 2.5 ms for 3D,  0.5 ms for $7\,E_r$ and 0.2 ms for 15-75 $E_r$.} }
    \label{Resonance_shift}
\end{figure}

We characterize the $\ket{1}-\ket{1}$ $p$-wave \del{Feshbach resonance}\add{FR} in 3D and quasi-1D by measuring atom loss as functions of $B$ and $\tau$. In 3D, we find the onset of loss at \del{$B_{3D}=$}159.05(1) G, which agrees with previous measurements of the location of this resonance in 3D \cite{Ketterle2005, Vale2008} but differs with other measurements \cite{Mukaiyama2013, Weidemuller2019} by a few 10's of mG. We are not able to resolve the expected doublet feature arising from the dipole-dipole interaction \cite{Ticknor2004, Weidemuller2019, Gunter2005} because of limitations of the field stability. \add{All the 1D data in this paper were measured with the magnetic field aligned with the z-axis, and thus only involve collisions with the $m_l=0$ projection of the angular momentum.} As $V_L$ is increased, we observe a confinement-induced shift in the resonance field and broadening of the atom-loss feature, as shown in Fig. \del{2}\add{\ref{Resonance_shift}}(a). 

We review $p$-wave scattering in 3D and quasi-1D to show how the measured confinement-induced shift can be used to extract $\alpha_p$, the 3D effective range. For low-energy collisions in 3D, the cotangent of the phase-shift $\delta_p$ associated with $p$-wave scattering can be expanded as a function of \del{two parameters, the }scattering volume, $V_p$, and \del{an }effective range, $\alpha_p$ \cite{joachain1975}:
\begin{equation}\label{eqn:cotdelta}
k^3 \cot(\delta_p (k))=-\frac{1}{V_p}-\alpha_p k^2 + O(k^4),
\end{equation}
\noindent where $\alpha_p>0$ and has units of inverse length. These scattering properties are modified in quasi-1D,
\begin{equation}
    k \cot(\delta_p(k))=-\frac{1}{l_p}-\xi_p k^2+ O(k^4),
\end{equation}
\noindent where $l_p$ is the 1D scattering length and $\xi_p$ is the 1D effective range, which has units of length. These quasi-1D scattering parameters are given by $l_p=3a_\perp\left[a_\perp^3 / \add{2}V_p+\alpha_p a_\perp+\mdel{3\sqrt{2}}\add{6}|\zeta(-1/2)|\right]^{-1}$ and $\xi_p=\alpha_p a_\perp^2/\mdel{3}\add{6}$ \cite{Granger2004,Pricoupenko2008,Shlyapnikov2017}, where $\zeta$ is the Riemann zeta function ($\zeta(-1/2) \approx -0.208$). The second and third terms in \del{$l_p$}\add{$1/l_p$} lead to a confinement-induced shift in the resonance location. \add{In this formalism, only dynamics along the axial dimension are relevant, and scattering quantities, such as the elastic scattering cross-section, are expressed in units appropriate for 1D.}

By performing a coupled-channel calculation, which requires detailed knowledge of the inter-atomic potentials \cite{Houbiers1998}, we obtain an expansion $1/V_p(B)$ up to second order in $B$. The effective range $\alpha_p$ can be approximated as a constant independent of $B$ for the relevant range of magnetic field. The \del{Feshbach resonance}\add{FR} in 3D occurs at the magnetic field $B_\mathrm{3D}$ at which $V_p$ diverges. Similarly, in quasi-1D, the resonance occurs when $l_p$ diverges at a magnetic field $B_\mathrm{1D}$, which is a function of $V_L$ and $\alpha_p$. The confinement-induced shift, $\delta_B(V_L,\alpha_p)=B_{1D}-B_{3D}$, can be approximated to leading order in confinement strength \add{$V_L$} by \del{(see Supplemental Material }\cite{SM}\del{)}
\begin{equation}
    \delta_B= \frac{-2 m E_r}{\hbar^2 \frac{\partial(1/V_p)}{\partial B}|_{B=B_{3D}}} \alpha_p \sqrt{V_L}.
    \label{eq:shift}
\end{equation}
\noindent \add{We cannot accurately measure $B_{3D}$ for $m_l=0$ alone due to the unresolved $|m_l|=1$ collisions in 3D, so} \del{W}\add{w}e fit the measured $\delta_B$ as a function of $V_L$ to Eq. \ref{eq:shift} by taking $\alpha_p$ and $B_{3D}$ as fitting parameters. The result of the fit to the quasi-1D data is shown by the solid curve in Fig. \del{2}\add{\ref{Resonance_shift}}(b). We obtain $\alpha_p=$ 0.14(1) $a_0^{-1}$ which is consistent with our coupled-channel result of 0.1412 $a_0^{-1}$, where $a_0$ is the Bohr radius\add{, and $B_{3D}=159.07(1)$ which is consistent with our loss-onset measurement and a dipolar splitting of 10 mG in 3D \cite{Weidemuller2019}}. We also find a consistent value by analyzing previous measurements performed on a \del{3}\add{2}D gas of $^6$Li in state $\ket{1}$ \cite{Mukaiyama2016,SM}.

\begin{figure}[b!]
    \centering
    \includegraphics[width=0.48\textwidth]{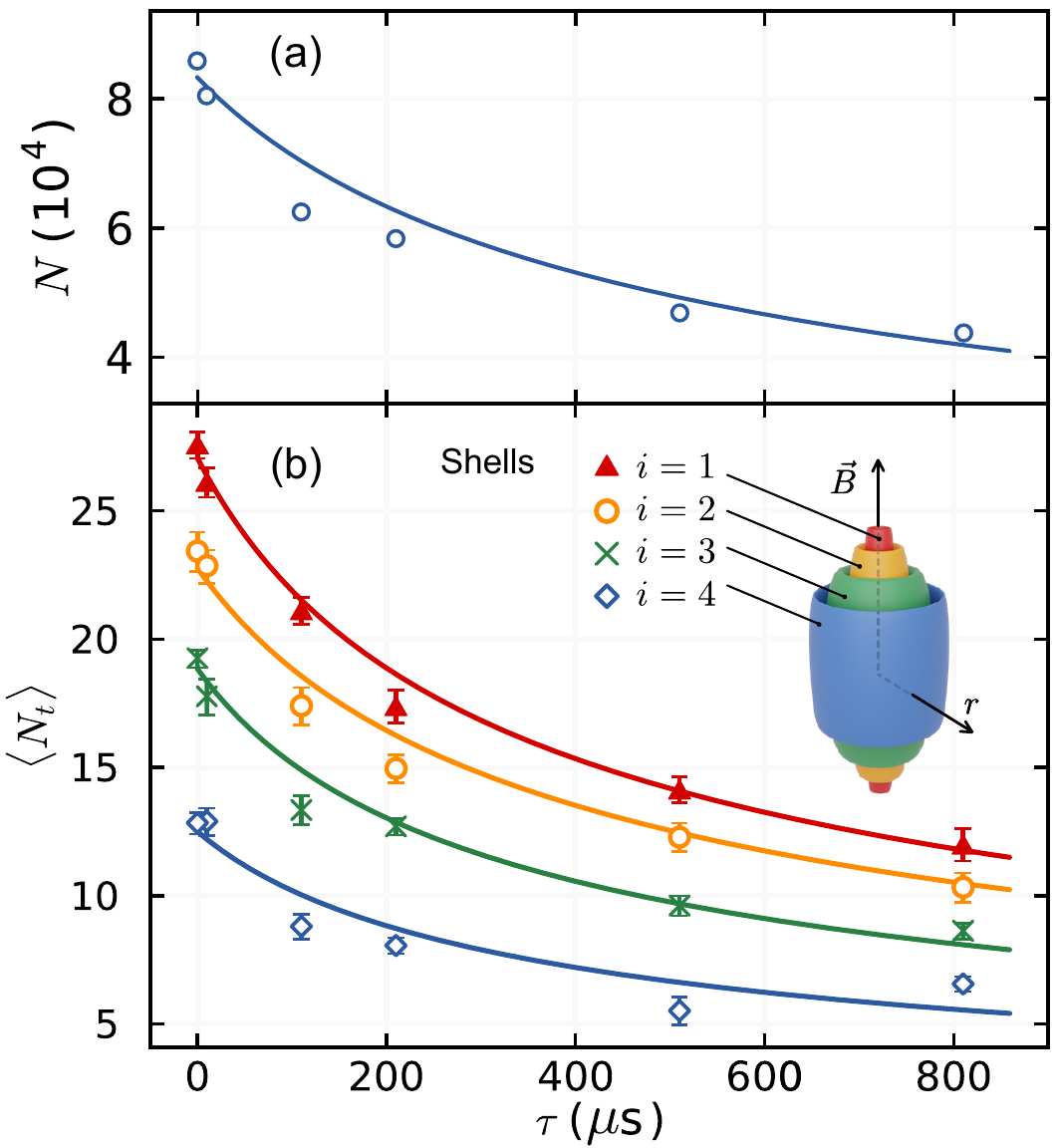}
    \caption{\del{FIG. 3}Typical time evolution of \add{(a) total number in the entire sample and (b)} averaged tube population $\langle N_t \rangle$ in 4 shells\add{.} \del{at}\add{For these data,} $\Delta B = $ 30 mG \del{with}\add{and} $V_L=$ 75 $E_r$. The different colors and symbols \add{in (b)} indicate different shells with approximately uniform initial atom number per tube. The shells are labeled from $i=1$, the inner-most, to $i=4$, the outer-most\del{ shell}. Solid curves show fits to Eq. \ref{eq:loss} to extract $L_3$ with the squared atomic density (a) $n^2=(N_{t,c}/2R_{F,c})^2$ of a central tube and (b) $n^2=(\langle N_{t} \rangle_i/2R_{F,i})^2$ of a typical tube in each shell. The corresponding $L_3$ values are plotted in Fig. \del{4}\add{\ref{Kad}}. Data points are averaged over 5 shots, and \del{error bars are }the standard error of the mean \add{is (a) approximately equal to the symbol size and (b) indicated by the error bars}.}   
    \label{loss_curve}
\end{figure}

The observed atom loss is presumably due to the formation of deeply-bound molecules. To characterize the \del{observed }loss, we measured $N$, the number of atoms remaining in the trap after a hold time $\tau$ for various $B$ and $V_L$. Background-gas collisions lead to a $1/e$ atom lifetime of 38 s in this apparatus, and are negligible for this analysis. Atom loss due to three-body collisions is described by 
\begin{equation}
    \frac{\dot{N}}{N}=-L_3\, n^2,
    \label{eq:loss}
\end{equation}
\noindent where $n^2=(N_{t,c}/2R_{F,c})^2$ is the squared atomic line density for a central tube, determined using a length-scale of twice the local Fermi radius $R_{F,c}$. We measure the time evolution with $V_L$ between 15 \del{to}\add{and} 75 $E_r$ and extract $L_3$ by fitting loss vs $\tau$ to Eq. \ref{eq:loss}. \add{Fig. \ref{loss_curve}(a) shows such a fit to typical loss data.} Since $L_3$ also depends on $\Delta B$, the field
detuning from resonance, we extract $L_3$ from the time evolution at several $\Delta B$ to find the peak value for each $V_L$. The peak $L_3$ for all $V_L$ are found to be approximately $7(2)\times10^{-6}$ cm$^2$/s. We observe no dependence on 1D confinement in this range \add{\cite{SM}}. Due to the inhomogeneity of the initial distribution of atoms across the 2D lattice, however, we find a rather poor agreement of the data to Eq. \ref{eq:loss}\del{ [40]}.

\begin{figure}[t!]
    \centering
    \includegraphics[width=0.48\textwidth]{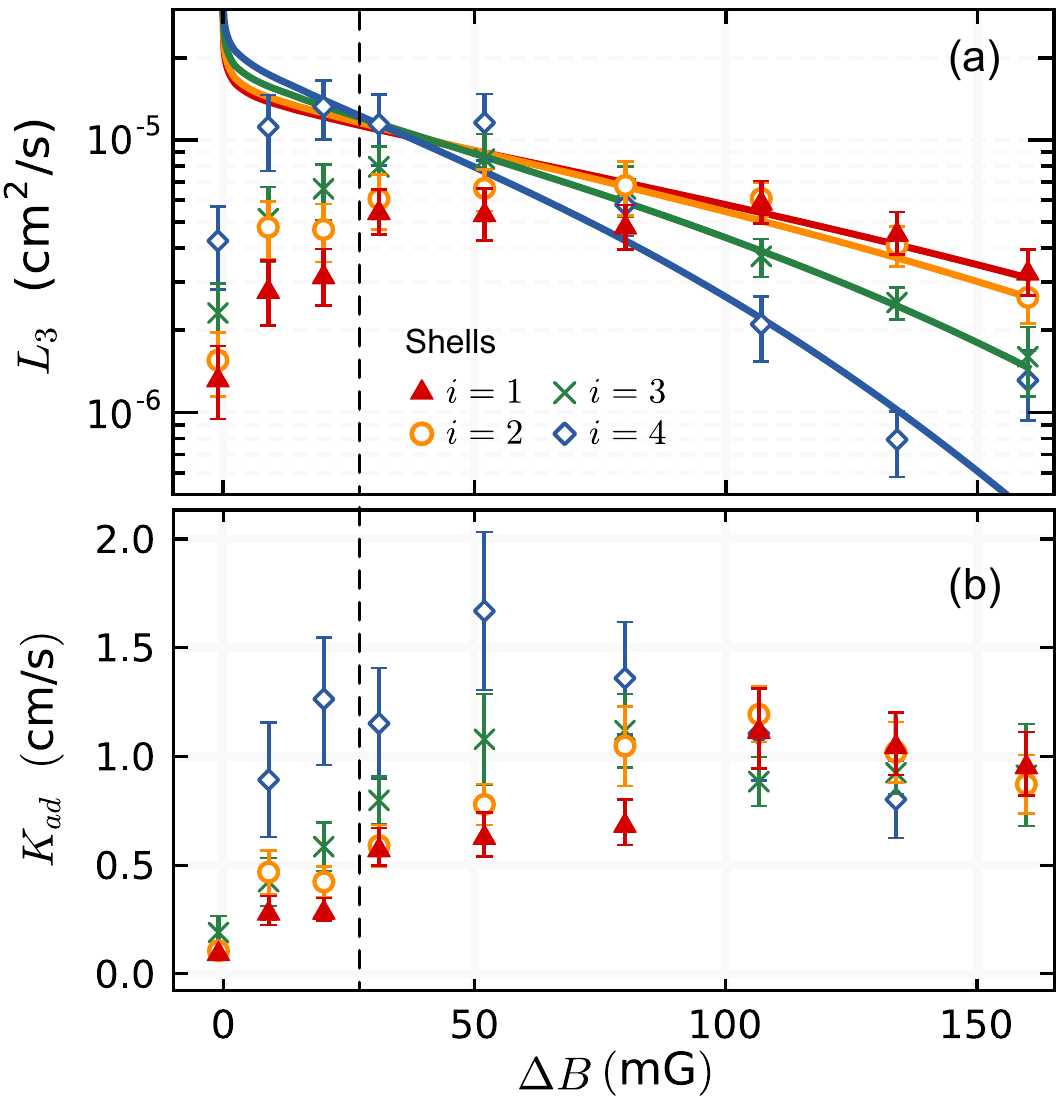}
    \caption {\del{FIG. 4}(a) $L_3$ \del{as a function of}\add{vs} $\Delta B$ for $V_L=$ 75 $E_r$. $L_3$ is \del{extracted from}\add{obtained by} fitting $N_{t,i}$ \add{vs $\tau$ to Eq. \ref{eq:loss}} for each shell\del{ vs $\tau$ with Eq. 4, as shown for}\add{. An example of this data is given in Fig. \ref{loss_curve}(b) for} $\Delta B=30$ mG\del{ in Fig. 3}. Solid curves show $(3/2)\hbar K_{ad}K_{aa}/\Gamma$ with a constant $\mdel{3}K_{ad}\mdel{/2}=$ \del{1}\add{0.67} cm/s, calculated for $T=0.1\ T_{F,1}$\add{, where $T_{F,1}=4.8(2)\ \mu$K}. (b) $K_{ad}$ \del{as a function of}\add{vs} $\Delta B$. $K_{ad}$ is extracted by fitting $\langle N_{t} \rangle_i$ vs $\tau$ to Eq. \ref{eq: NaNd}, using the calculated values of $\Gamma$ and $K_{aa}$. Black dashed line indicates $\Delta B=$ 27 mG, which corresponds to $\kappa a_\perp=$ 1/2 for $V_L=$ 75 $E_r$ \cite{Cui2017}. Error bars \del{indicate the}\add{are} one-sigma confidence interval\add{s} for the fitting parameters $L_3$ and $K_{ad}$\del{, respectively}. \add{The large uncertainty in the fitted values for the outermost shell is indicative of small $N_t$.}}  
    \label{Kad}
    
\end{figure}

The results of \del{an alternative}\add{a more comprehensive} analysis of the same data that provides an improved fit \add{to Eq. \ref{eq:loss}} is shown in Fig. \del{3}\add{\ref{loss_curve}(b)}. Here, we group the tubes into separate cylindrical shells (labeled by $i=1-4$) with an averaged atom number per tube $\langle N_t \rangle_i$ \add{\cite{SM}} and a corresponding Fermi temperature $T_{F,i}$. Figure \del{4}\add{\ref{Kad}}(a) shows $L_3$ for each shell extracted from data with $V_L=75$ $E_r$ \del{and $\Delta B=30$ mG}\add{vs $\Delta B$}. The peak $L_3$ for each shell is in the range of $5\times10^{-6}$ cm$^2$/s to $1\times10^{-5}$ cm$^2$/s, and is similar to the peak $L_3$ extracted from the whole atomic cloud.

\del{Recently, the loss of ultracold fermions near a $p$-wave Feshbach resonance in 3D has been modeled as a cascade of two consecutive two-body processes. Two atoms resonantly form a Feshbach dimer, followed by a collision between the dimer and an atom, resulting in a deeply-bound molecule and an atom [26, 27]. The differential equations governing this loss process are}\add{In \cite{Cui2017}, Zhou and Cui suggest that the rate of three-body loss near a $p$-wave FR can be suppressed by reducing the overlap between the wavefunctions of a deeply-bound molecule and a Feshbach dimer with increasing confinement. To investigate this hypothesis, we analyze our observed loss data using a cascade model of two consecutive two-body processes instead of a direct three-body event: two atoms resonantly form a dimer, followed by a collision between the dimer and an atom, resulting in a deeply-bound
molecule and an atom \cite{BoGao2018}. This approach has previously been applied to the particular $p$-wave FR we study, but in 3D and quasi-2D \cite{Mukaiyama2019}. It is the natural formalism in which to evaluate the predicted suppression, as it models the formation and relaxation of dimers. The equations governing this loss process are}

\begin{subequations}
\begin{align}
    \frac{dN_a}{dt} &=2\frac{\Gamma}{\hbar}N_d-2 K_{aa} \frac{N_a(N_a-1)}{4R_F}-K_{ad}\frac{N_aN_d}{2R_F},\tag{5\add{a}}\\
    \frac{dN_d}{dt} &=-\frac{\Gamma}{\hbar}N_d+K_{aa}\frac{N_a(N_a-1)}{4R_F}-K_{ad}\frac{N_aN_d}{2R_F},\tag{\del{6}\add{5b}}
\end{align}
\label{eq: NaNd}
\end{subequations}

\noindent where $N_a$ is the number of atoms, $N_d$ is the number of \del{Feshbach }dimers, $K_{aa}$ is the two-body event rate for atom-atom collisions converting atoms into \del{Feshbach }dimers, and $K_{ad}$ is the two-body atom-dimer inelastic collision event rate. \add{$\Gamma$, t}\del{T}he \add{one-body} decay rate of dimers is\del{ determined by} the width of the \del{Feshbach resonance}\add{FR}\del{, $\Gamma$}. The rate of dimer formation is proportional to the number of possible pairs of atoms, given by $N_a(N_a-1)/2!$\del{ and typically approximated as $N_a^2/2$ for large $N_a$. However, in our experiment, the atom number per quasi-1D tube is quite small ($\approx 30$), and thus we use the exact form in our analysis}.

\add{$K_{ad}$ is of particular interest, as it depends on the overlap between dimers and deeply-bound molecules.} Both $\Gamma$ and $K_{aa}$ are related to the elastic scattering cross-section, $\sigma_{1D}(E)$, which can be calculated, thus constraining the fit to the cascade process \del{by}\add{to} a single \del{fitting }parameter, $K_{ad}$. $\sigma_{1D}(E)$ may be approximated by a Lorentzian in collision energy, $E=\hbar^2k^2/m$\add{, centered at the above-threshold binding energy of the Feshbach dimer $E_\mathrm{res}=-\hbar^2/ l_p\xi_p m>0$ and with width $\Gamma=(\hbar/\xi_p)\sqrt{4E_\mathrm{res}/m}$} \del{as follows }[\add{\citenum{Gurarie2007},} \citenum{SM}]\del{:}\add{.}
\del{\noindent Here, $E_\mathrm{res}=-\hbar^2/ l_p\xi_p m>0$ for $l_p<0$ is the above-threshold binding energy of the Feshbach molecule, and $\Gamma=(\hbar/\xi_p)\sqrt{4E_\mathrm{res}/m}$ [6].}

\del{The scattering cross-section may be used to calculate $K_{aa}$, which is}\add{$K_{aa}$ may be calculated by averaging}\del{ the average of} $\sigma_\mathrm{1D}(k_r)$ over the ensemble of pairs of atoms with relative momentum $k_r$ and velocity $v_r$
\begin{equation}\label{eqn:Kaa}
    K_{aa}=\langle\sigma_\mathrm{1D}(k_r)v_r\rangle=\hbar \int^\infty_{-\infty} dk_r\,\sigma_\mathrm{1D}(k_r)v_rP(k_r),\tag{\del{8}\add{6}}
\end{equation}
\noindent where $P(k_r)$ is the probability density function of $k_r$ obtained from the density distribution of a trapped Fermi gas \cite{SM}. We assume a global temperature $T$ across the entire sample. However, $\mu$ varies significantly from tube-to-tube due to the density inhomogeneity across the 2D lattice. This effect is mitigated by sectoring the cloud into shells of similar $\mu$, as discussed earlier, thus giving a distinct value of $K_{aa}$ for each shell. For each quasi-1D tube, $\mu$ is determined by $N_{t,j}$ and $T$.

Although we cannot directly measure $T$, we exploit the fact that at a sufficiently large $\Delta B$, the rate equations can be approximated \del{by}\add{as} a \add{direct} three-body loss \del{equation}\add{process} with \add{a loss coefficient }$\mdel{L_3}\add{\Tilde{L_3}} = (3/2\mdel{!}) \,\hbar K_{ad}K_{aa}/\Gamma$ \add{under the assumptions of a steady-state dimer population ($dN_d/dt=0$) and $\Gamma/\hbar \gg K_{ad} N_a / 2R_F$} \cite{Mukaiyama2019}. \add{Assuming that these assumptions hold for large $\Delta B$, we fit the measured values of $L_3$ for each shell with $T$ and $K_{ad}$ as fitting parameters to $\Tilde{L_3}$.} \del{By comparing $\hbar K_{aa}/\Gamma$ to our measured values of $L_3$ for $\Delta B>$ 100 mG, as shown in Fig. 4(a), we determine}\add{We find that} $T = 0.1\, T_{F,1}$, and \del{by carrying out the integrations in Eq. 6 numerically, we obtain $K_{aa}$ for each shell. }\add{that $K_{ad}=0.67$ cm/s is independent of field for $\Delta B>$ 100 mG. The assumptions given above are confirmed in this range. The solid lines in Fig. \ref{Kad}(a) show $\Tilde{L_3}$ for each shell.}

The extracted $K_{ad}$ \add{values from fitting loss data} for $V_L=$ 75 $E_r$ \add{to Eqs. \ref{eq: NaNd} using the calculated values of $\Gamma$ and $K_{aa}$} \del{is}\add{are} shown in Fig. \del{4}\add{\ref{Kad}}(b) \add{for the full range of $\Delta B$ \cite{SM}}. \del{The relative uniformity of $K_{ad}$ for $\Delta B>50$ mG has previously been cited as evidence that the cascade model is appropriate and conforms to a previous analysis of 3D and quasi-2D measurements on the same Feshbach resonance [27]. The large uncertainty in the fitted value for the outermost shell is indicative of its small $N_t$.}\add{We find that under these conditions, Eqs. \ref{eq: NaNd} model the time behavior of the observed loss as well as Eq. \ref{eq:loss}. The values of $K_{ad}$
extracted for $\Delta B>50$ mG are field independent. The observed field independence strongly supports the cascade model as the atom-dimer collision process is inherently non-resonant. In the dimer formation step, the atoms must collide with a momentum dictated by the binding energy of the dimer, which is field-dependent. The dimer relaxation step, however, may proceed for any collision momentum, as the atom receives the binding energy of the deeply bound molecule.}


The behavior of $K_{ad}$ for $\Delta B<50$ mG \del{could be a consequence of the suppression predicted by Zhou and Cui [25], which would manifest as a suppression of $K_{ad}$. The stretching of the quasi-1D $p$-wave molecule wavefunction is expected to be significant for $\sqrt{2}\kappa a_\perp <1/2$, where $\kappa=\sqrt{m E_\mathrm{res}}/\hbar$ [25]. This predicted suppression is strongest for small $\Delta B$, where $E_\mathrm{res}$ is smallest.}\add{is consistent with a suppression of the rate of dimer relaxation. The spatial overlap of the dimer and deeply-bound wavefunctions increases with $\kappa a_\perp$, where $\kappa=\sqrt{mE_\mathrm{res}}/\hbar$, so the predicted suppression is strongest for small $\Delta B$, where $E_\mathrm{res}$ is smallest. The suppression is expected to be significant for $\kappa a_\perp <1/2$ \cite{Cui2017}},\del{.} \add{which f}\del{F}or $V_L=$ 75 $E_r$\del{, this condition} corresponds to $\Delta B<$ 27 mG. \del{The suppression we observe is relatively mild (around a factor of 2), however, and only relevant in a narrow magnetic field range near the resonance, limiting its experimental usefulness. }Another interpretation of the small-detuning behavior of $K_{ad}$ is that the cascade model breaks down\del{ and is no longer applicable, perhaps due to unitarity effects [24], or} due to\add{, for example,} the existence of a shallow three-body bound state\del{ as recently suggested by Schmidt \textit{et al.}} \cite{Schmidt2019}.

This work is the first detailed experimental study of $p$-wave collisions in quasi-1D. We confirm the confinement-induced shift and broadening as a function of $V_L$. The confinement-induced shift agrees well with quasi-1D theory \cite{Shlyapnikov2017} and the extracted value \del{for the 3D effective range}\add{of} $\alpha_p$ agrees with previous work \cite{Mukaiyama2016}. We measure $L_3$ as a function of $V_L$ and find no dependence up to 75 $E_r$. The magnetic field independence of $K_{ad}$ for $\Delta B>50$ mG confirms the cascade model \cite{BoGao2018,Mukaiyama2019} for three-body loss in quasi-1D in the regime of large \del{magnetic field detunings}\add{$\Delta B$ ($>100$ mG)}, as well as for intermediate \del{magnetic field detunings}\add{$\Delta B$ (50-100 mG)} where \del{the loss cannot be}\add{the cascade model is not well} approximated by the three-body loss \add{rate} equation.

The suppression in $K_{ad}$ at $\Delta B<50$ mG is possibly explained by $p$-wave \del{molecule}\add{dimer} stretching \cite{Cui2017}. Achieving greater suppression \add{in $^6$Li} by increasing $V_L$ \del{would be}\add{is} challenging since at a fixed $\Delta B$, $\kappa a_\perp \propto 1/V_L^{1/4}$ \cite{SM}\del{. Although we do not have clear evidence of a suppression of atom loss, future work at even higher $V_L$ could be fruitful as the increased confinement should expand the magnetic field range over which the molecular wavefunction is extended. Similarly, improved magnetic field resolution and stability would enable further study of this narrow feature closer to resonance. These technical advancements could improve the prospects of detecting $p$-wave pairing and emulating the Kitaev chain Hamiltonian.}\add{, but future work at even higher $V_L$ or with improved magnetic field resolution and stability would enable further study of this narrow feature. Our result also provides insight into a potential pathway towards observing pairing between identical fermions in cold atom systems. Suppressing loss in heavier fermions with FRs, such as $^{40}$K \cite{Regal2003,Gunter2005,Thywissen2016}, $^{161}$Dy \cite{Lev2014}, and $^{167}$Er \cite{Ferlaino2018}, is promising, as small values of $\kappa a_\perp$ may be more readily achieved in these atoms.}

\add{\textit{Note added.} $-$ During the peer-review process, another group reported on a similar experiment \cite{ohara2020}. Although both groups observe similar overall atom loss, they report a suppression of $L_3 \propto V_L^{-1}$, while we find $L_3$ independent of $V_L$ over a wide range (Fig. S2). The difference lies in the choice between defining $L_3$ using the 3D or the 1D densities. In their analysis, $L_3$ is defined in terms of the 3D density of a tube, which increases with $V_L^{1/2}$, while we use the 1D line density.  While the two results are consistent, we argue that 1D densities are most appropriate based on physical and practical considerations.  Physically, the dimensionless quantity $\kappa a_{\perp}$ parameterizes the effective dimensionality of the system near a FR, and the peak values of $L_3$ we report were measured in regions where $\kappa a_{\perp}<1$. Practically, 1D units make it clear that the peak loss rate is independent of $V_L$.}

We would like to thank T. L. Yang for his contributions to the apparatus and W. I. McAlexander for his coupled-channel code. This work was supported in part by the Army Research Office Multidisciplinary University Research Initiative (Grant Nos. W911NF-14-1-0003 and W911NF-17-1- 0323), the NSF (Grant No. PHY-1707992), and the Welch Foundation (Grant No. C-1133). D. C. acknowledges financial support from CONACyT (Mexico, Scholarship No. 472271). 

\bibliography{main}

\end{document}


\title{Supplementary material for ``Collisional loss of one-dimensional fermions near a $p-$wave Feshbach resonance''}


\author{Ya-Ting Chang}
    \affiliation{Department of Physics and Astronomy, Rice University, Houston, Texas 77005, USA}

\author{Ruwan Senaratne}
    \affiliation{Department of Physics and Astronomy, Rice University, Houston, Texas 77005, USA}
    
\author{Danyel Cavazos-Cavazos}
    \affiliation{Department of Physics and Astronomy, Rice University, Houston, Texas 77005, USA}
    
\author{Randall G. Hulet}
    \email[ ]{randy@rice.edu}
    \affiliation{Department of Physics and Astronomy, Rice University, Houston, Texas 77005, USA}    
\date{\today} 

\maketitle
\renewcommand{\theequation}{S\arabic{equation}}
\renewcommand{\thefigure}{S\arabic{figure}}

\section*{Confinement-induced shift $\delta_B$ in quasi-1D}

The Feshbach resonance occurs at a magnetic field $B_{1D}$ where the 1D $p$-wave scattering length $l_p$ diverges \cite{Granger2004,Pricoupenko2008,Shlyapnikov2017}

\begin{equation}
    \frac{1}{l_p}=\frac{a_\perp^3/\add{2}V_p+\alpha_p a_\perp+\mdel{3\sqrt{2}}\add{6}|\zeta(-1/2)|}{\add{3a_\perp}} = 0.
    \label{eq:divergence}
\end{equation}
\noindent Since $\alpha_p a_\perp \gg \mdel{3\sqrt{2}}\add{6}|\zeta(-1/2)|$ for the lattice depths $V_L$ we can achieve in this experiment, Eq. \ref{eq:divergence} can be approximated using $1/V_p=-\add{2}\alpha_p/a^2_{\perp}$. By Taylor expanding this around the 3D resonance field $B_{3D}$ to first order
\begin{equation}
    \frac{1}{V_p}|_{B=B_{1D}}=\frac{1}{V_p}|_{B=B_{3D}} + \frac{\partial(1/V_p)}{\partial B}|_{B=B_{3D}} (B_{1D}-B_{3D}) = -\frac{\add{2}\alpha_p}{a_{\perp}^2},
    \label{eq:taylor_Vp}
\end{equation}
\noindent we obtain a simple analytical form for the confinement-induecd shift $\delta_B(V_L,\alpha_p)=B_{1D}-B_{3D}$
\begin{align}
    \delta_B&= \frac{-\add{2}\alpha_p}{\frac{\partial(1/V_p)}{\partial B}|_{B=B_{3D}} a_{\perp}^2} \nonumber \\
    &=\frac{-2 m E_r}{\hbar^2 \frac{\partial(1/V_p)}{\partial B}|_{B=B_{3D}}} \alpha_p \sqrt{V_L},
    \label{eq:shift_final}
\end{align}
\noindent where $\frac{\partial(1/V_p)}{\partial B}|_{B=B_{3D}} < 0$.

\section*{Confinement-induced shift $\delta_{B,2D}$ in quasi-2D}

\begin{figure}[!h]
    \centering
    \includegraphics[width=0.51\textwidth]{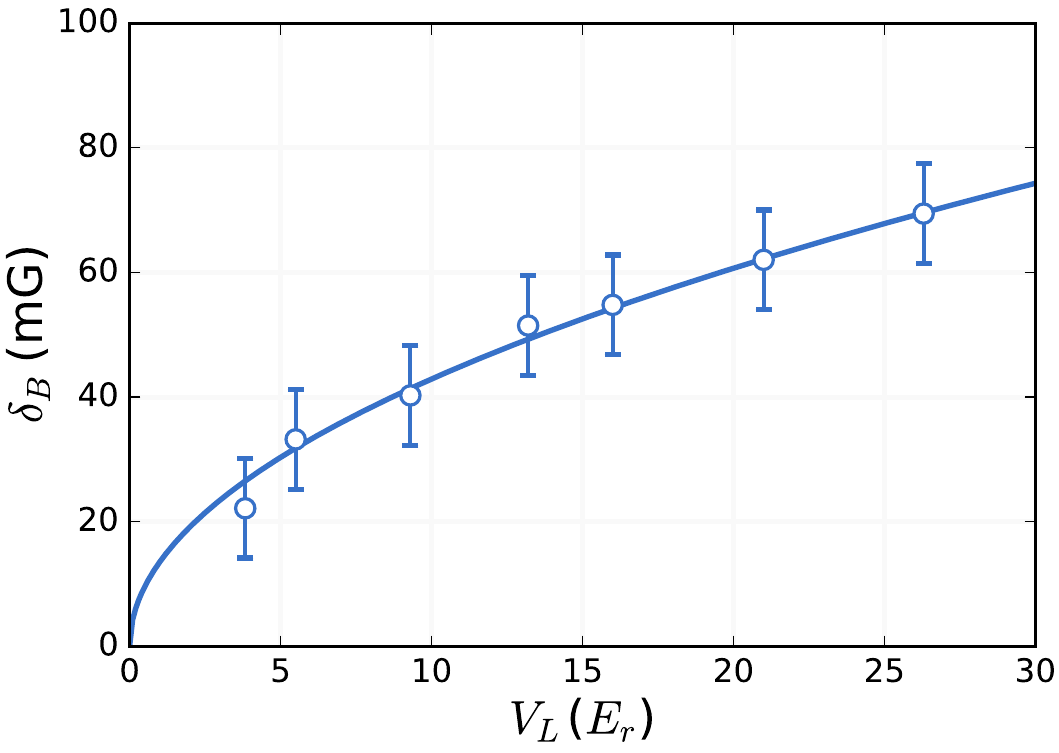}
    \caption{The open circles show the confinement-induced shift for the quasi-2D $p$-wave Feshbach resonance position as a function of the corresponding lattice depth (data from \cite{Mukaiyama2016}). The solid curve shows the result of fit to Eq. \ref{eq:shift}, with the effective range $\alpha_p = $ 0.158(5) $a_0^{-1}$.}
    \label{Resonance_shift_Q2D}
\end{figure}

Similar\add{ly} to the confinement-induced shift in quasi-1D, an equivalent expression to Eq. \ref{eq:shift_final} can be derived for this geometry by considering the quasi-2D scattering parameters \cite{Shlyapnikov2017}. The confinement-induced shift in quasi-2D $\delta_{B,2D}=B_{2D}-B_{3D}$ can be approximated by
\begin{equation}
    \delta_{B,2D}=\frac{1}{2}\delta_{B}.
    \label{eq:shift}
\end{equation}
The open circles in Fig. \ref{Resonance_shift_Q2D} show\del{s} the data of $\delta_B$ in quasi-2D from \cite{Mukaiyama2016}, and the solid curve shows the result of a fit to Eq. \ref{eq:shift} with the effective range $\alpha_p = $ 0.158(5) $a_0^{-1}$ as a fitting parameter. This value of $\alpha_p$ is within 15\% of the value the authors of \cite{Mukaiyama2016} obtained by fitting measurements of the dissociation energy, as well as the value extracted from the fit to our quasi-1D data shown in the main text.

\section*{Three-body loss coefficient $L_3$ versus $V_L$}
\begin{figure}[!bh]
    \centering
    \includegraphics[width=0.5\textwidth]{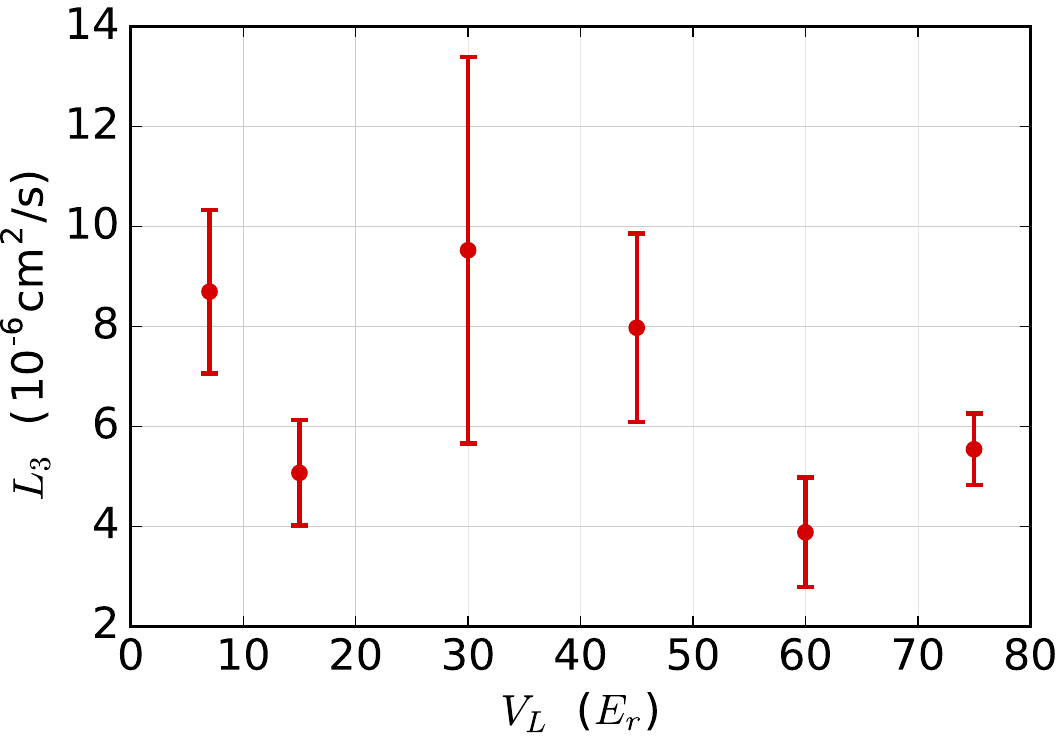}
    \caption{\del{Figure S3. }Peak three-body loss coefficient $L_3$ as a function of lattice depth $V_L$. $L_3$ is extracted by fitting the time evolution of $N$ to Eq. \del{S5}\add{4} as described in the Fig. \del{S2}\add{2(a)} caption. Error bars indicate the one-sigma confidence interval for the fitting parameters $L_3$.}   
    \label{L3_VL}
\end{figure}

\del{Atom loss due to three-body collisions is governed by}
\del{\noindent The three-body loss coefficient $L_3$ is obtained by fitting the time evolution of the total number of atoms $N$ to Eq. S5, where $n^2=(N_{t,c}/2R_{F,c})^2$ is the squared atomic line density for a central tube, with a volume of twice the local Fermi radius $R_{F,c}$. A typical time evolution and fit to Eq. S5 is shown in Fig. S2.} 
We measure the time evolution \add{of atom number $N$} with $V_L$ between 15 \del{to}\add{and} 75 $E_r$ and extract $L_3$ by fitting to Eq. \del{S5}\add{4 in the main text}. The peak $L_3$ as a function of $V_L$ is shown in Fig. \del{S3}\add{\ref{L3_VL}}.

\section*{Analysis using \textit{in situ} imaging with inverse Abel transform}
\begin{figure}[!h]
    \centering
    \includegraphics[width=1.0\textwidth]{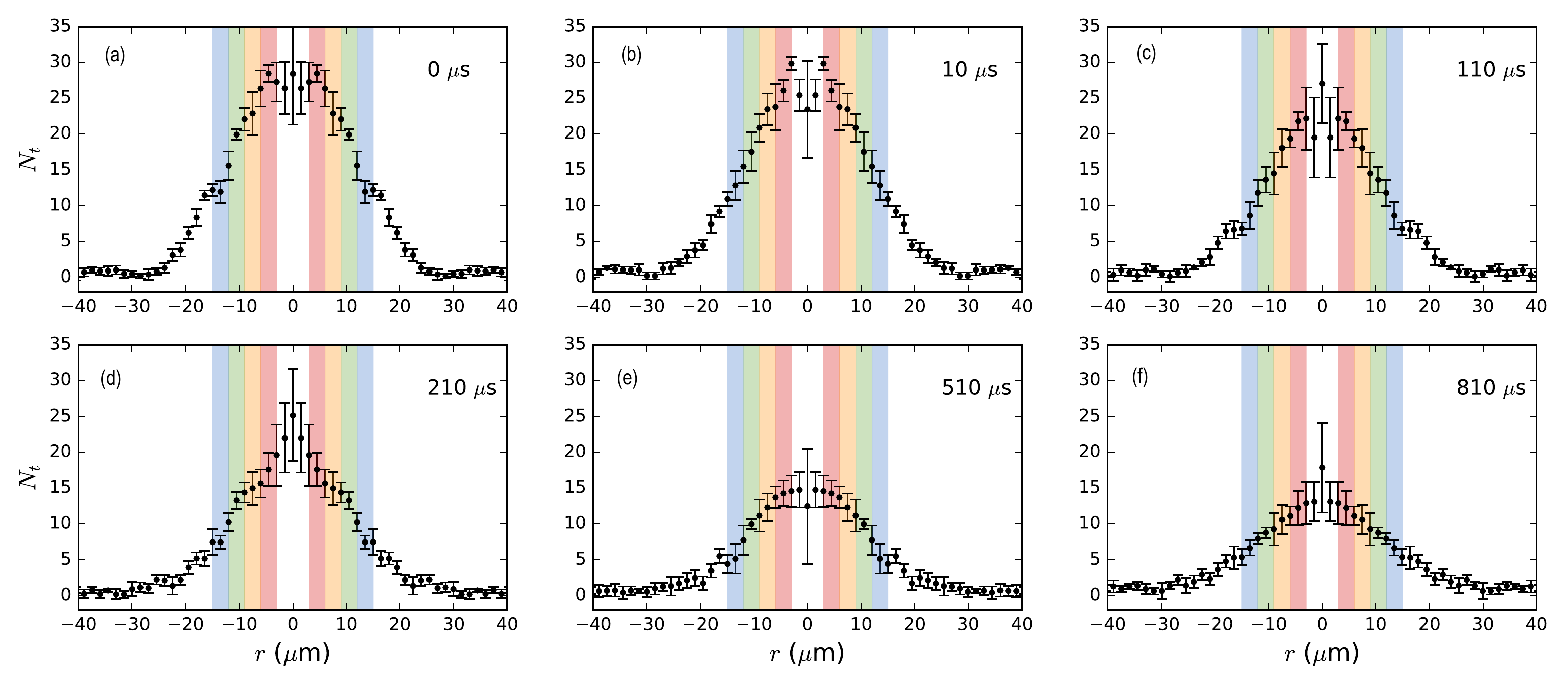}
    \caption{\add{Typical time evolution of $N_t$ in a tube at a distance $r$ from the center of the 2D lattice. Regions with colored backgrounds correspond to the shells $i=1-4$ in the main text Fig. 2(b). Data points are averaged over 5 shots, and error bars indicate the standard error of the mean.}}   
    \label{insitu}
\end{figure}

\add{We probe atoms using \textit{in situ} imaging and perform the inverse Abel transform on the column density, assuming cylindrical symmetry, to obtain the distribution of the number of atoms per tube $N_t(r)$. We group the tubes into separate cylindrical shells as shown in Fig. \ref{insitu}.}
\section*{Typical time evolution of $\langle N_t \rangle$ fits to the cascade model }

\begin{figure}[!h]
    \centering
    \includegraphics[width=0.52\textwidth]{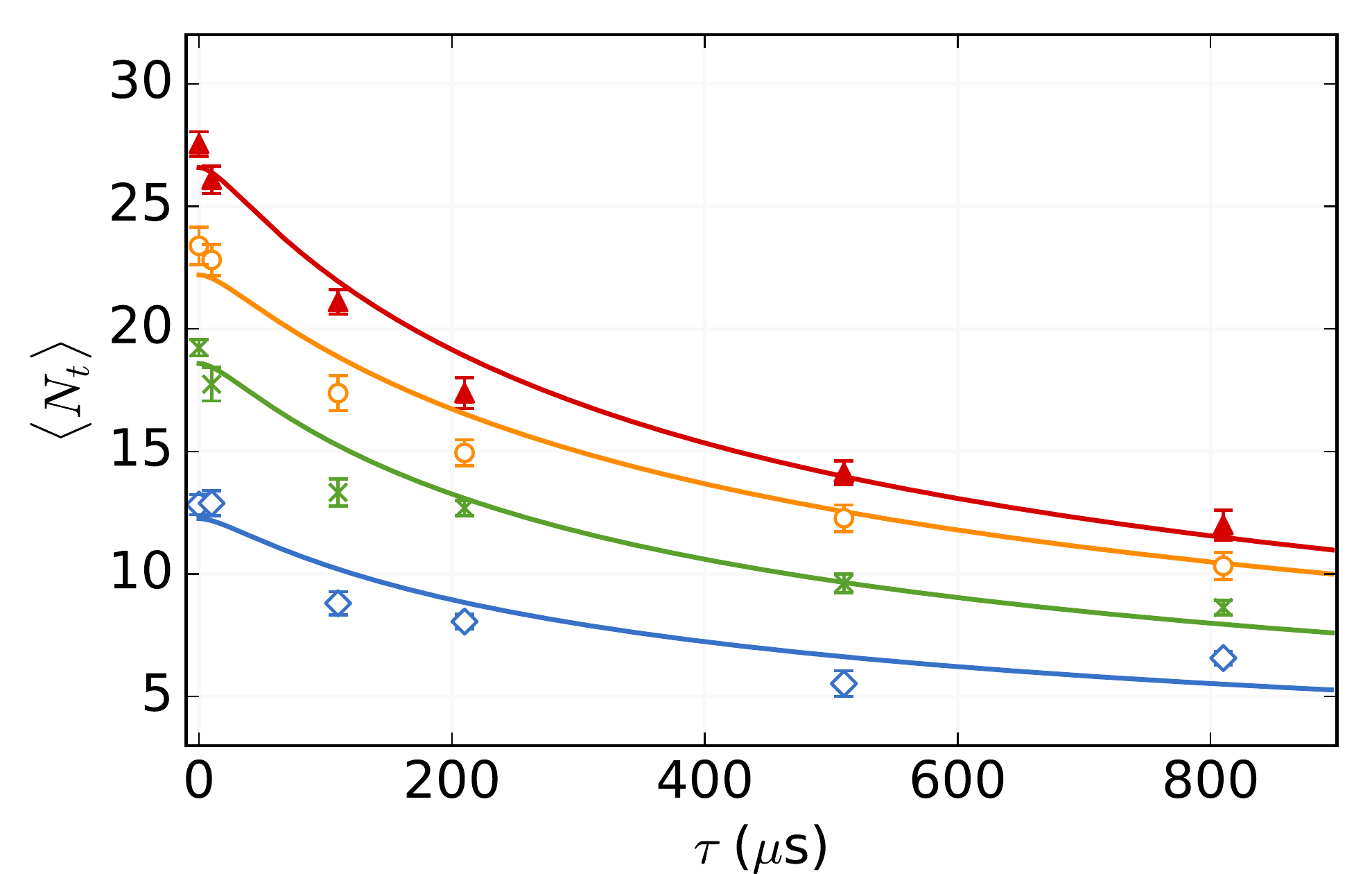}
    \caption{\add{Typical time evolution of averaged tube population $\langle N_t \rangle$ in 4 shells at $\Delta B = $ 30 mG with $V_L=$ 75 $E_r$. Data are the same as Fig. 2(b) in the main text. Solid curves show fits to Eq. 5 in the main text with $K_{aa}$ and $\Gamma$ calculated for $T=0.1\ T_{F,1}$.} }  
    \label{indirect}
\end{figure}

\add{We extract $K_{ad}$ from the fits to Eq. 5 in the main text, using theoretical values of $K_{aa}$ and $\Gamma$ for $T=0.1\ T_{F,1}$. A typical time evolution with the fitted curve is shown in Fig. \ref{indirect}.}

\section*{Quasi-1D $p$-wave scattering cross-section}

The quasi-1D $p$-wave scattering amplitude is \cite{Shlyapnikov2017}
\begin{equation}
    f_\mathrm{1D}(k)=\frac{-ik}{1/l_p+\xi_pk^2+ik},
    \tag{S\del{6}\add{5}}
\end{equation}

\noindent In 1D, the equivalent of the scattering cross-section is simply the modulus of the scattering amplitude squared
\begin{equation}
    \sigma_\mathrm{1D}(k)=|f_{1D}(k)|^2=\frac{k^2}{k^2+(1/l_p+\xi_pk^2)^2},
    \tag{S\del{7}\add{6}}
\end{equation}
\noindent which is bounded from above by 1. Near a Feshbach resonance for $l_p<0$, this expression may be approximated by a Lorentzian in terms of the collision energy, $E=\hbar^2k^2/m$ as follows: 
\begin{equation}\label{eqn: sigmaapprox}
    \sigma_\mathrm{1D}(E)\approx\frac{\left(\frac{\Gamma}{2}\right)^2}{(E-E_\mathrm{res})^2+\big(\frac{\Gamma}{2}\big)^2}.
    \tag{S\del{8}\add{7}}
\end{equation}

\noindent Here, $E_\mathrm{res}=-\hbar^2/ l_p\xi_p m>0$ for $l_p<0$ is the above-threshold binding energy of the Feshbach molecule, and $\Gamma=(\hbar/\xi_p)\sqrt{4E_\mathrm{res}/m}$ is the width of the resonance \cite{Gurarie2007}.

\section*{Probability density function of $k_r$ for a trapped Fermi gas}

The probability density function of $k_r$ for a trapped Fermi gas is
\begin{equation}\label{eqn:convolution}
    P(k_r)=\frac{\hbar}{N_a^2}\int^\infty_{-\infty} dk' n(k')n(k'-k_r), 
    \tag{S\del{9}\add{8}}
\end{equation}

\noindent and the $k$-space number density is given by

\begin{equation}\label{eqn:momdist}
    n(k)=\frac{1}{2\pi\hbar}\int^\infty_{-\infty} dx \frac{1}{\mathrm{exp}[\beta(\frac{1}{2}m\omega_\mathrm{z}^2x^2+\frac{\hbar^2k^2}{2m}-\mu)]+1},
    \tag{S\del{10}\add{9}}
\end{equation}

\noindent where $\beta=1/k_\mathrm{B}T$ and $\mu$ is the chemical potential.

\section*{ Dependence of $\kappa a_\perp$ on $V_L$ }
$\kappa$ is the magnitude of the wavevector related to the binding energy of the Feshbach molecule, which can be calculated by
\begin{equation}
    \kappa= \frac{\sqrt{mE_\mathrm{res}}}{\hbar}=\sqrt{\frac{-1}{l_p \xi_p}}.
    \label{eq:kappa}
    \tag{S\del{11}\add{10}}
\end{equation}
\noindent where $l_p$ and $\xi_p$ are the 1D scattering parameters modified from 3D scattering quantities $V_p$ and $\alpha_p$ with confinement strength $a_\perp$ as mentioned in the main text. By Taylor expanding $\kappa^2$ around the Feshbach resonance field $B_{\mathrm{1D}}$ for a particular $V_L$, we find a constant $\kappa$ at a fixed magnetic field detuning $\Delta B$ which is independent of $V_L$:

\begin{align}
    \kappa^2(\Delta B) &= \kappa^2\vert_{B=B_{1D}} + \frac{\partial \kappa^2}{\partial B}\vert_{B=B_{1D}}\Delta B + \frac{\partial^2 \kappa^2}{\partial B^2}\vert_{B=B_{1D}}\Delta B^2 \add{+ O(\Delta B^3)} \nonumber \\
    &=\frac{\partial (1/V_p)}{\partial B}\vert_{B=B_{1D}}\Delta B + \frac{\partial^2 (1/V_p)}{\partial B^2}\vert_{B=B_{1D}}\Delta B^2 \add{+O(\Delta B^3)}
    \label{eq:kappa_2}
    \tag{S\del{12}\add{11}}
\end{align}

\begin{figure}[!h]
    \centering
    \includegraphics[]{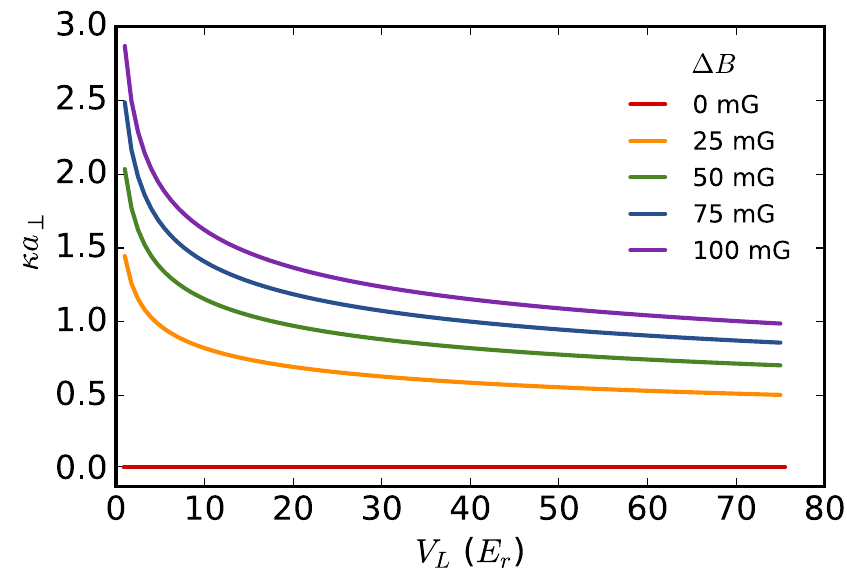}
    \caption{\del{Figure S4. }$\mdel{\sqrt{2}}\kappa a_\perp$ as a function of lattice depth $V_L$. Each curve is $V_L^{1/4}$ with a scaling factor set by $\Delta B$.}   
    \label{kappa_aperp}
\end{figure}

Therefore, $\mdel{\sqrt{2}}\kappa a_{\perp}$ is proportional to $V_L^{1/4}$ for a particular $\Delta B$, as shown in Fig. \del{S4}\add{\ref{kappa_aperp}}.

\bibliography{main}